\documentclass{elsart}
\journal{WORP2 Workshop \date{March 11, 2004}}
\usepackage{amsmath}
\usepackage{amsfonts}
\usepackage{amssymb}
\usepackage{natbib}
\usepackage{subfigure}
\bibpunct{[}{]}{;}{a}{}{;}	
\usepackage[dvips]{graphicx}
\usepackage{url}

\def\woprfigs{}

\begin{document} 
\begin{frontmatter}

\title{Benchmarking Blunders and Things That Go Bump in the Night}
\author{Neil J. Gunther}
\address{Performance Dynamics Company,\\
4061 East Castro Valley Blvd., Suite 110,
Castro Valley, CA 94552, USA} 
\thanks{Copyright~\copyright ~2004 Performance Dynamics Company. All Rights Reserved.}
\ead{njgunther@perfdynamics.com}
\ead[url]{www.perfdynamics.com}

\begin{abstract}
Benchmarking---by which I mean any computer system that is driven by a
controlled workload---is the ultimate in performance simulation. Aside from
being a form of institutionalized cheating, it also offer countless
opportunities for systematic mistakes in the way the workloads are applied
and the resulting measurements interpreted. Right test, wrong conclusion is
a ubiquitous mistake that happens because test engineers tend to treat data
as divine. Such reverence is not only misplaced, it's also a sure ticket to
production hell when the application finally goes live. I demonstrate how
such mistakes can be avoided by means of two war stories that are real
\emph{WOPR}s.
(a) How to resolve benchmark flaws over the psychic hotline and
(b) How benchmarks can go flat with too much Java juice! 
In each case I present simple performance models and show how they
can be applied to correctly assess benchmark data.
\end{abstract}


\end{frontmatter}

\section{Introduction}
Benchmarking is the ultimate in performance analysis (i.e., workload simulation).
It is often made more difficult because it takes place in a competitive
environment: be it vendors competing against each other publicly using the
TPC \url{www.tpc.org}, or SPEC \url{www.spec.org} benchmarks~\citep[See
e.g.,][]{benchbook} or a customer assessing vendor performance running their
own application during the procurement cycle. And a lot happens in the
stealth of night.

Notwithstanding the fact that benchmarking is a form of institutionalized
cheating---which everybody knows but won't admit publicly---there are
countless opportunities for blunders in the way the loads are constructed
and applied. Everybody tends to focus on that aspect of designing and
running benchmarks.

A far more significant problem---and one that everyone seems to be blissfully
unaware of---is interpreting benchmark data correctly. Right test, wrong
conclusion is a much more common blunder than many of us realize. I submit
that this happens because test engineers tend to treat performance data as divine.
The huge cost and effort required to set up a benchmark system can lead us
all into the false security that the more complex the system, the more
sacrosanct it is. As a consequence, whatever data it generates must be
correct by design and to even suspect otherwise is sacrilegious. Such
reverence for performance test data is not only misplaced, it often
guarantees a free trip to hell when the application finally goes live.

\begin{figure}[!hbtp]
\centering
\includegraphics[bb = 0 0 322 218, scale = 1.0]{\woprfigs 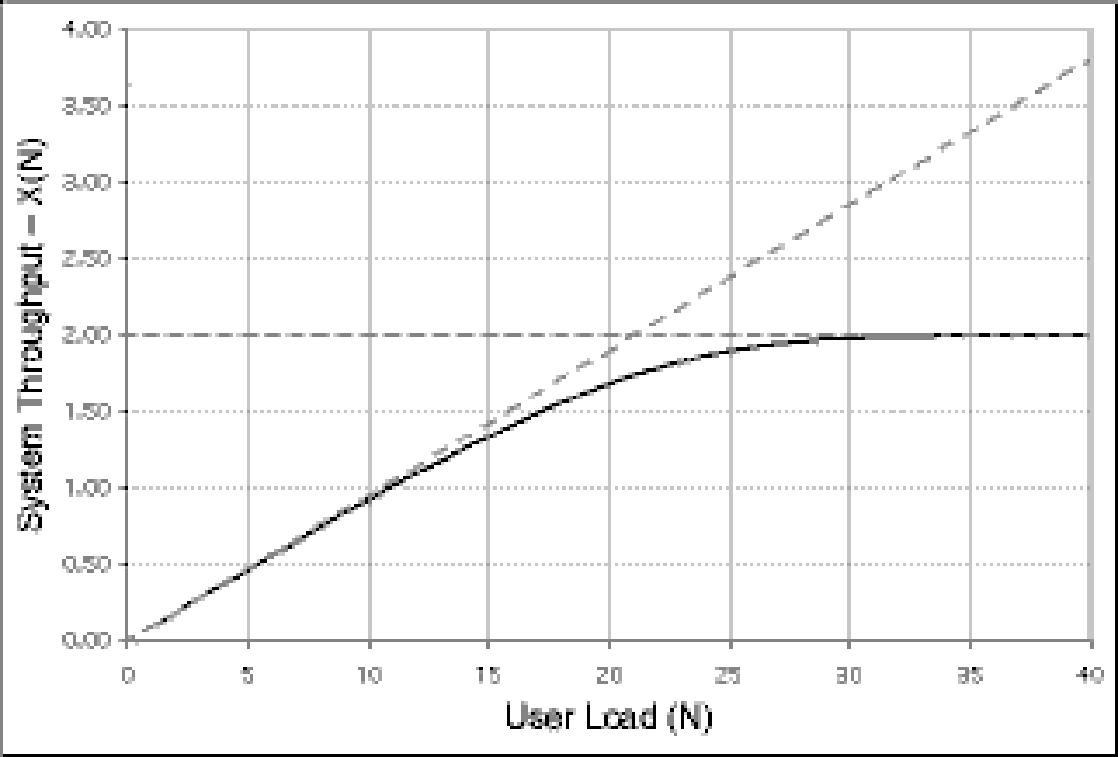} 
\caption{Canonical throughput characteristic.} 
\label{fig:convexX}
\end{figure}

In this paper, I demonstrate by example how such benchmark blunders arise
and, more importantly, how they can be avoided with simple performance
models that provide a correct conceptual framework in which to interpret the 
data.

\begin{figure}[!hbtp]
\centering
\includegraphics[width=5.5in]{\woprfigs 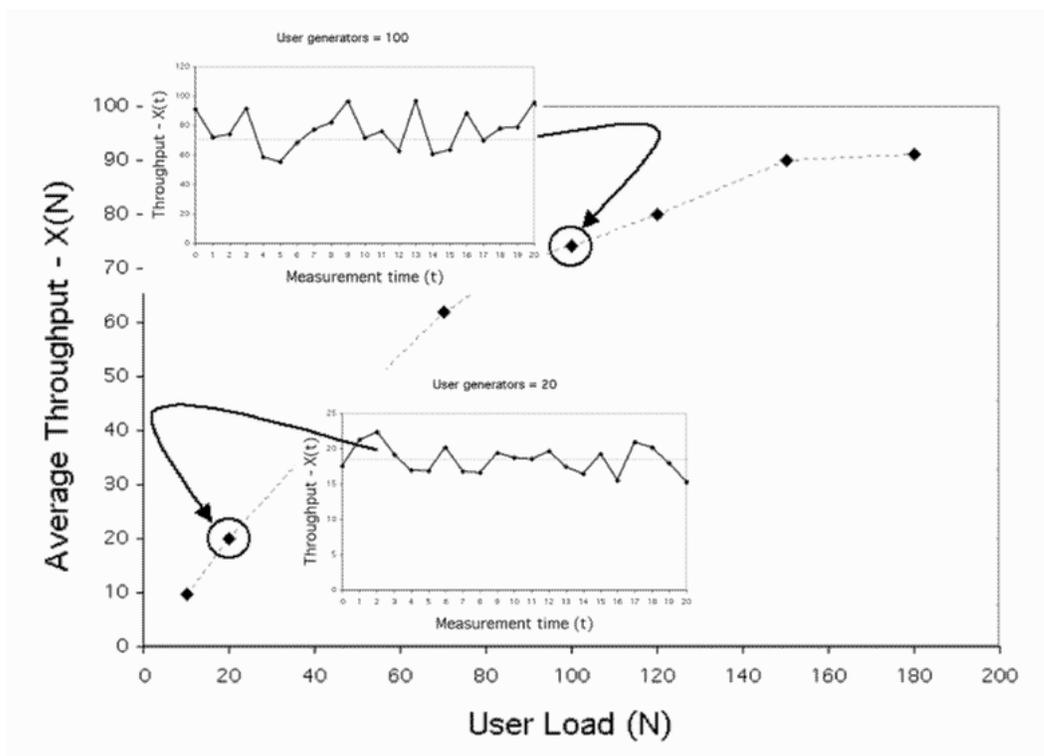} 
\caption{Relationship between steady-state measurements of the
instantaneous throughput $X(t)$ (insets) and the time-averaged throughput
$X(N)$ (dots).}
\label{fig:steadystate}
\end{figure}

Fig.~\ref{fig:convexX} shows the canonical system throughput characteristic
$X$ (the dark curve). This curve is generated by taking the statistical
average of the instantaneous throughput measurements at successive client
load points $N$ once the system has reached \emph{steady
state}~\citep[cf.][]{Willenborg} (as d.psed in Fig.~\ref{fig:steadystate}).

\section{Canonical Curves} \label{sec:canonical}
In the subsequent discussion, I shall refer to some canonical performance
characteristics that occur in all benchmark measurements.

The dashed lines in Fig.~\ref{fig:convexX} represent bounds on the
throughput characteristic. The horizontal dashed line is the ceiling on the
achievable throughput $X_{max}$. This ceiling is controlled by the
bottleneck resource in the system; which also has the longest service time
$S_{max}$. The variables are related inversely by the formula:
\begin{equation}
X_{max} = \frac{1}{S_{max}} \label{eqn:xmax}
\end{equation}
which tells us that the bigger the bottleneck, the lower the maximum throughput;
which is why we worry about bottlenecks.
The point on the $N$-axis where the two bounding lines intersect is a first
order indicator of optimal load $N_{opt}$. In this case, $N_{opt} = 21$ VUsers.

The sloping dashed line in Fig.~\ref{fig:convexX} shows the best case
throughput if their were no contention for resources in the system---it
represents \emph{equal bang for the buck}---an ideal case that cannot be
achieved in reality.
\begin{figure}[!hbtp]
\centering
\includegraphics[bb = 0 0 322 218, scale = 1.0]{\woprfigs 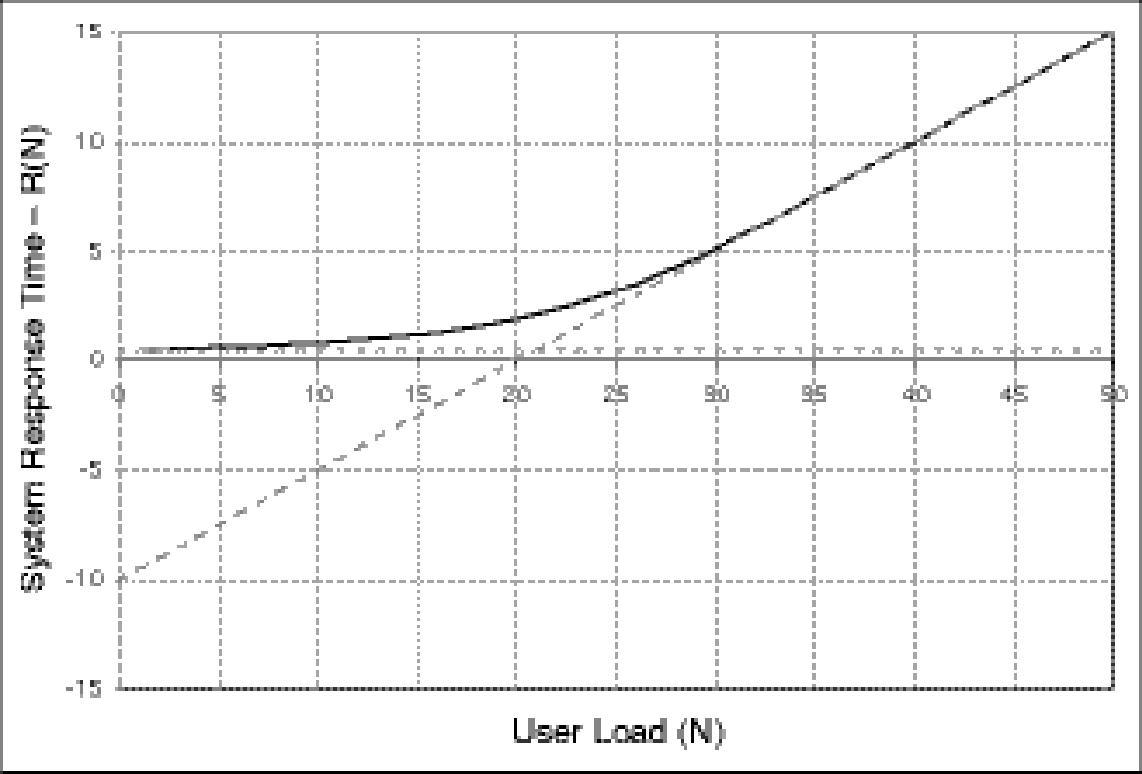} 
\caption{Canonical delay characteristic.} 
\label{fig:convexR}
\end{figure}

Similarly, Fig.~\ref{fig:convexR} shows the canonical system response time
characteristic $R$ (the dark curve). This shape is often referred to as the
response \emph{hockey stick}. It is the kind of curve that would be
generated by taking time-averaged delay measurements in \emph{steady state}
at successive client loads.

The dashed lines in Fig.~\ref{fig:convexR} also represent bounds on the
response time characteristic. The horizontal dashed line is the floor of
the achievable response time $R_{min}$. It represents the shortest possible
time for a request to get though the system in the absence of any
contention. The sloping dashed line shows the worst case response time once
saturation has set in.
These things will constitute our principal performance models.

\begin{figure}[!hbtp]
\centering
\includegraphics[bb = 0 0 669 454, scale = 0.5]{\woprfigs 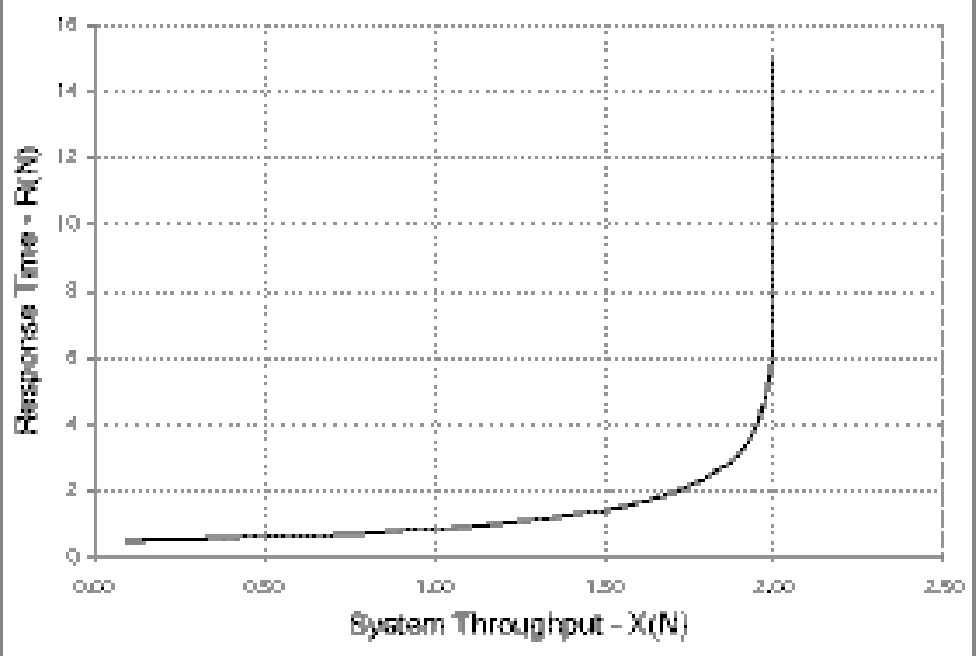} 
\caption{Non-canonical throughput-delay curve.}
\label{fig:qclosedXR}
\end{figure}

In passing, I note that representations of throughput $X(N)$ and 
response time $R(N)$ can be combined into a single plot like 
Fig.~\ref{fig:qclosedXR}~\citep[See e.g.,][]{Splaine}.
Although useful in some contexts, the combined plot suffers from the
limitation of not being able to calculate the location of $N_{opt}$.

\section{Psychic Hotline and Benchmark Mentalism} \label{sec:psychic}

\subsection{Background}
Dateline: Miami Florida, sometime in the mirky past. Names and numbers
have been changed to protect the guilty. I was based in San Jose. All
communications occurred over the phone. I never went to Florida and I
never saw the benchmark platform.

Over the prior 18 months, a large benchmarking rig had been set up to test
the functionality and performance of some third party software. Using this
platform, the test engineers had consistently measured a system throughput
of around 300 TPS (transactions per second) with an average think time of
10 s between sequential transaction requests. Moreover, during the
course of development, the test engineers had managed to have the
application instrumented so they could see where time was being spent
internally when the application was running. This is a good thing and a
precious rarity!

\subsection{Benchmark Results}
The instrumented application had logged internal processing times. In the
subsequent discussion we'll suppose that there were \emph{three} sequential
processing stages. Actually, there were many more. Enquiring about typical
processing times reported by this instrumentation, I was given a list of
average values which I shall represent by the three token values: $3.5,
5.0$, and $2.0$ ms. At that point I responded, ``Something is rotten in
Denmark ... err .... Florida!''

\subsection{The Psychic Insight} \label{sec:insight}
So, this is our first benchmarking blunder.
I know from the application instrumentation data that the bottleneck
process has a service time of $S_{max} = 0.005$ s
and applying the bounds analysis of Sect.~\ref{sec:canonical} I know:
\begin{equation}
X_{max} = \frac{1}{0.005} = 200 \; \textrm{TPS}
\end{equation}
I can also predict the optimal user load as:
\begin{equation}
N_{opt} = \frac{3.5 + 5.0 + 2.0 + (10.0 \times 1000)}{5} = 2002.1 \; \textrm{VUsers}
\end{equation}
where I have replaced the 10 s think time by 10,000 ms. But the same data
led the Florida test engineers to claim 300 TPS as their maximum throughput
performance. From this information I can hypothesize that either:
\begin{enumerate}
\item the benchmark measurement of 300 TPS is wrong or,
\item the instrumentation data are wrong.
\end{enumerate}
Once this inconsistency was pointed out to them, the test engineers
decided to thoroughly review their measurement methodology~\footnote{
Note that recognizing any inconsistency at all is half the battle.}. 
We got off the phone with the agreement that we would resume the discussion
the following day. During the night~\footnote{The other part of the title of
this paper.}, the other shoe dropped.

The client scripts contained an \texttt{if()} statement to calculate the
instantaneous think time between each transaction in such a way that its
statistical mean would be 10 s. The engineers discovered that this code
was not being executed correctly and the think time was effectively zero seconds.

In essence, it was as though the transaction measured at the client
$X_{client}$ was comprised of two contributions:
\begin{displaymath}
X_{client} = X_{actual} + X_{errors}
\end{displaymath}
such that $X_{actual} = 200$ TPS and $X_{errors} = 100$ TPS (except these
weren't bona fide transactions). In other words, the test platform was
being overdriven in \emph{batch} mode (zero think time) where the unduly
high intensity of arrivals caused the software to fail to complete some
transactions correctly. Nonetheless, the application returned an
\textsc{ack} to the benchmark driver which then scored it as a correctly
completed transaction. So, the instrumentation data were correct but the
benchmark measurements were wrong. This led to a performance claim for the
throughput that was only in error by 50\%!

Even after 18 months hard labor, the test engineers remained blissfully
unaware of this margin of error because they were not in possession of the
simple performance models presented in Sect.~\ref{sec:canonical}. Once they
were apprised of the inconsistency, however, they were very quick to assess
and correct the problem~\footnote{The more likely incentive is that they
urgently wanted to prove me wrong.}. The engineers in Florida did all the
work, I just thought about it. Dionne Warwick and her Psychic Friends
Network would've been proud of me.

\section{Falling Flat on Java Juice!}

\subsection{Background}
The following examples do not come directly from my own experience but
rather have been chosen from various books and reports. They represent a
kind of disturbing \emph{syndrome} that, once again, can only arise when
those producing the test data do not have the correct conceptual framework
within which to interpret it.

\subsection{Demonized Performance Measurements} \label{sec:httpd}
\begin{figure}[!hbtp]
\centering
\includegraphics[bb = 0 0 700 716, scale = 0.5]{\woprfigs 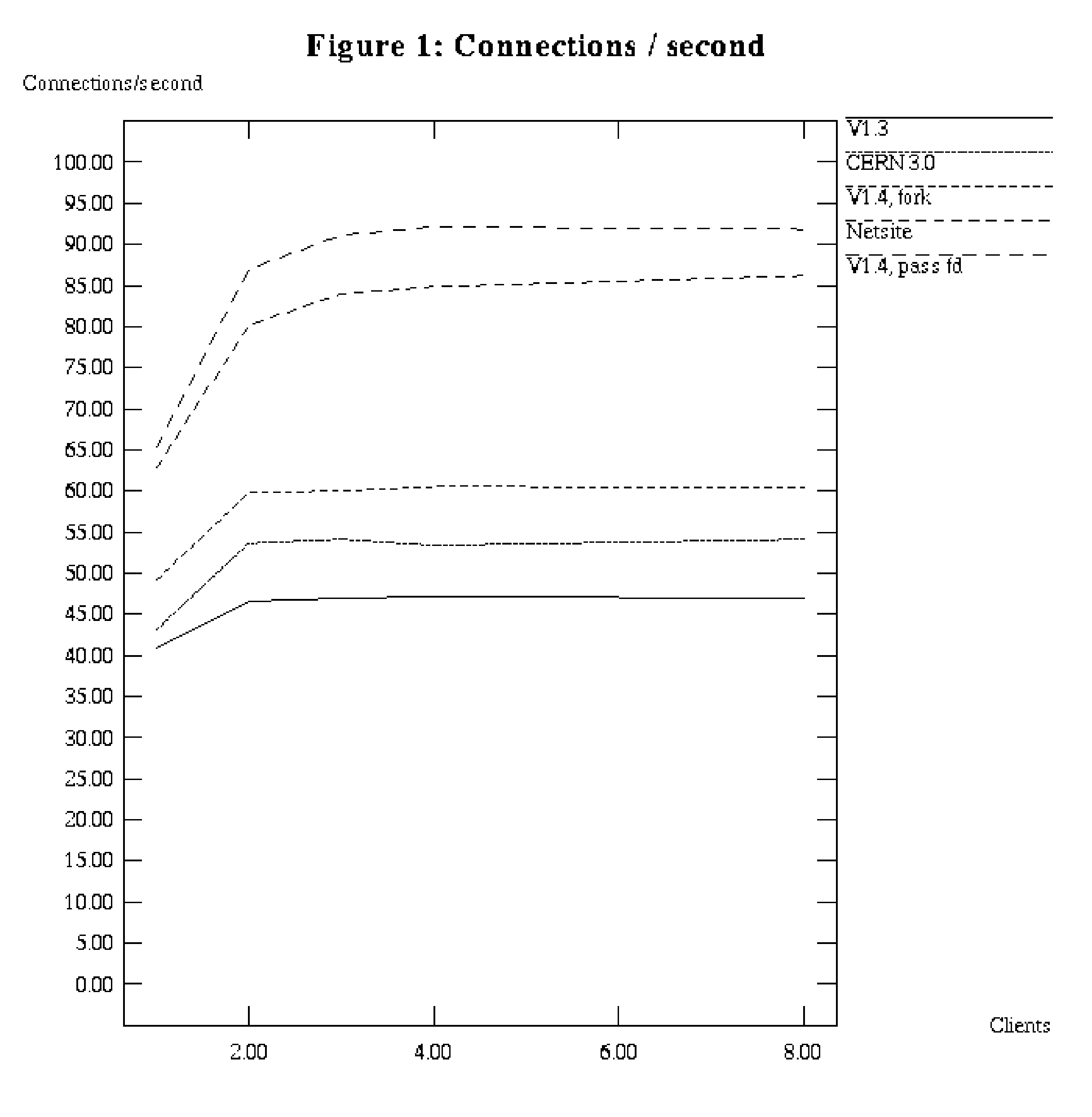} 
\caption{Measured throughput (conn / s) for a suite of HTTPd servers.} 
\label{fig:strangeNCSAX}
\end{figure}

\begin{figure}[!hbtp]
\centering
\includegraphics[bb = 0 0 700 716, scale = 0.5]{\woprfigs 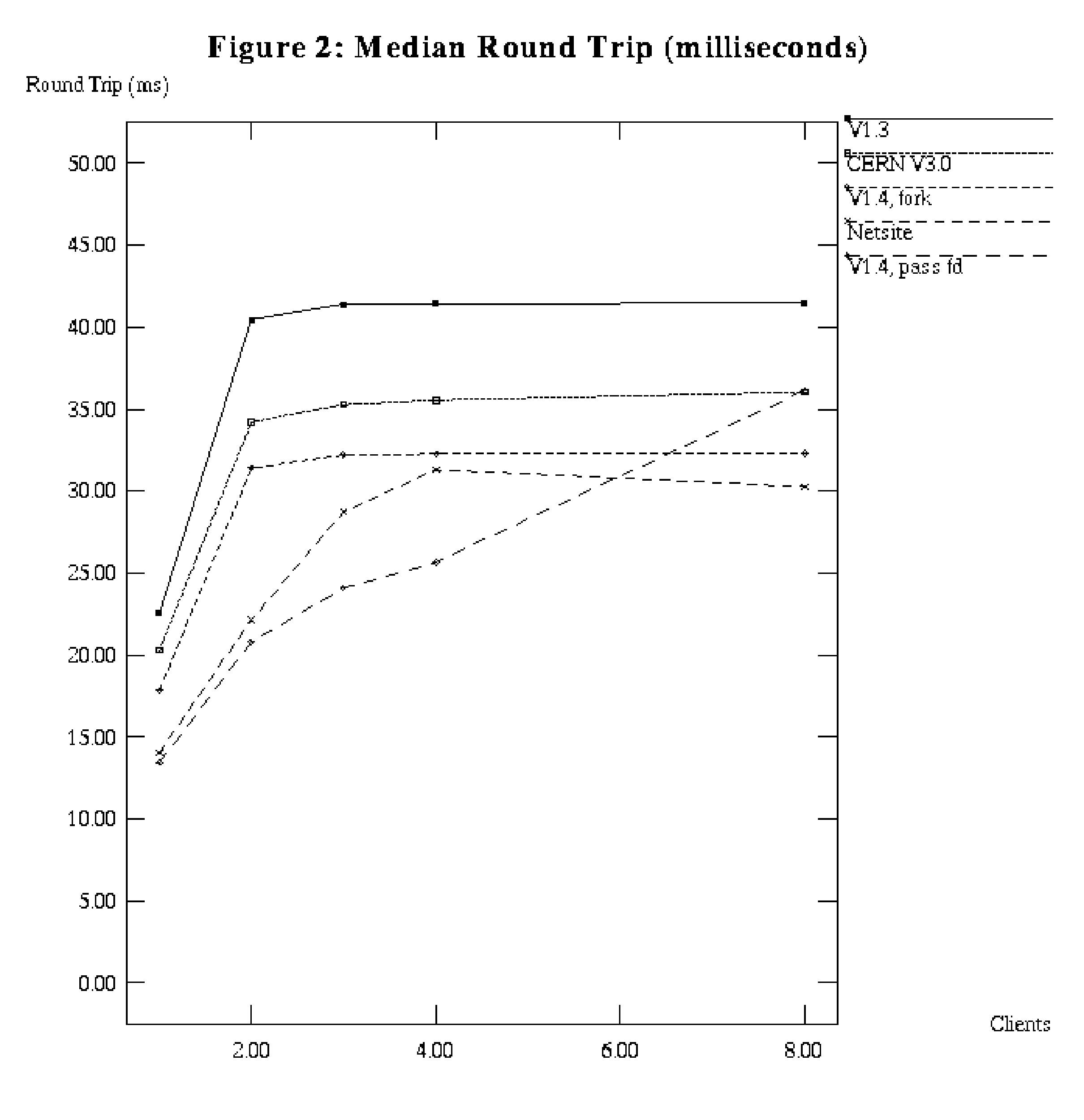} 
\caption{Measured response times (ms) of the same suite of HTTPd servers.} 
\label{fig:strangeNCSAR}
\end{figure}

The load test measurements in Fig.~\ref{fig:strangeNCSAX} were made on a
variety of HTTP demons~\citep{McGrath}.
Each curve roughly exhibits the expected throughput characteristic
as described in Sect.~\ref{sec:canonical}.
The slightly odd feature, in this case, is the fact that the servers under test
appear to saturate rapidly for user loads between 2 and 4 clients. Turning to
Fig.~\ref{fig:strangeNCSAR}, we see similar features in those
the curves.
Except for the bottom curve, the other four curves appear to
\emph{saturate} between $N = 2$ and 4 client generators. Beyond the knee
one dashed curve even exhibits \emph{retrograde} behavior; something that
would take us too far afield to discuss here.  The interested reader should
see~\citep[][Chap. 6]{njgBOOK00}.

But these are supposed to be \emph{response time} characteristics, 
not throughput characteristics. According to Sect.~\ref{sec:canonical} this
should never happen! This is our second benchmarking blunder. These
data defy the laws of queueing theory~\citep[][Chap. 2, 3]{njgBOOK00}. Above
saturation, the response time curves should start to climb up a
\emph{hockey stick} handle with a slope determined by the bottleneck 
service time $S_{max}$. But this is not an isolated mistake.

\subsection{Java Juiced Scalability} \label{sec:javaperf}
\begin{figure}[!th]
\begin{center}
  \subfigure[``Undesirable'' scalability.]{
    \label{fig:javaA}
     \includegraphics[bb = 0 0 322 218, scale = 0.5]{\woprfigs 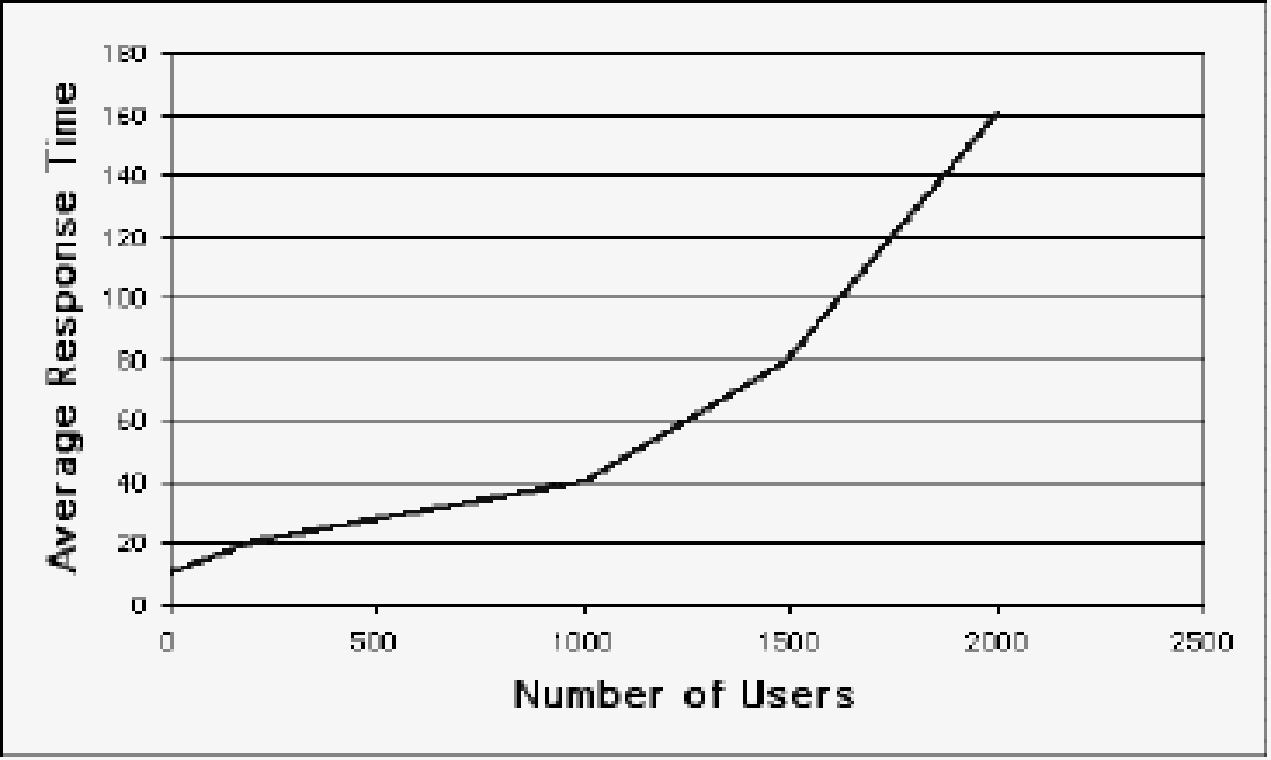} }     
	 \hskip 0.01\textwidth
     \subfigure[``Desirable'' scalability.]{
     \label{fig:javaB}
     \includegraphics[bb = 0 0 322 218, scale = 0.5]{\woprfigs 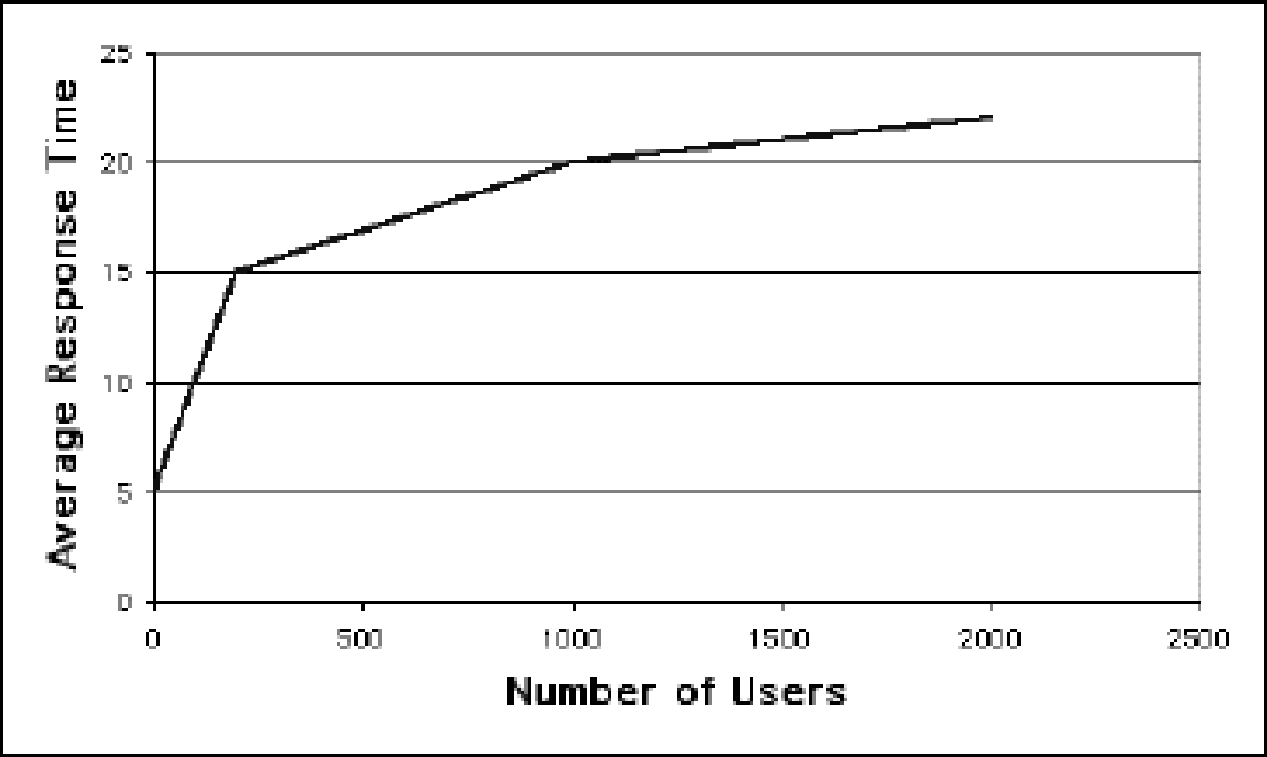}}
\end{center}
\caption{Comparison of java scalability taken from \citep{Wilson}.}
\label{fig:javacurves}
\end{figure}

In their book on Java performance analysis, \citet[][pp. 6-7]{Wilson}
refer to the classic convex response time characteristic in
Fig.~\ref{fig:convexR} of Sect.~\ref{sec:canonical} as being undesirable
for good scalability. I quote ...
\begin{quote}
\it (Fig.~\ref{fig:javaA}) ``isnÕt scaling well'' because response time
is increasing ``exponentially'' with increasing user load.
\end{quote}

They define \emph{scalability} rather narrowly as ``the study of how
systems perform under heavy loads.'' This is not necessarily so. As I have
just shown in Sect.~\ref{sec:httpd}, saturation may set in with just a few
active users. Their conclusion is apparently keyed off the incorrect
statement that the response time is increasing ``exponentially'' with
increasing user load. No other evidence is provided to support this claim.

Not only is the response time \emph{not} rising exponentially, the application may
be scaling as well as it can on that platform. I know from
Sect.~\ref{sec:canonical}, that saturation is to be expected and growth above 
saturation is \emph{linear}, not exponential. Moreover, such behavior does not, by
itself, imply poor scalability. Indeed, the response
time curve may even rise \emph{super-linearly} in the presence of thrashing
effects but this special case is not discussed either.

These authors then go on to claim that a response time characteristic of the
type shown in Fig.~\ref{fig:strangeNCSAR} is more desirable.
\begin{quote}
\it (Fig.~\ref{fig:javaB}) ``scales in a more desirable manner''
because response time degradation is more gradual with increasing user
load.
\end{quote}
Assuming these authors did not mislabel their own plots (and their text
indicates that they did not), they have failed to comprehend that the
\emph{flattening} effect is a signal that something is wrong. Whatever the
precise cause, this \emph{snapping} of the hockey stick handle should be
interpreted as a need to scrutinize their measurements, rather than hold them
up as a gold standard for scalability.

This is not a benchmarking blunder per se but, as the physicist Wolfgang
Pauli once retorted, ``This analysis is so bad, it's not even wrong!''

\subsection{Threads That Throttle} 
\begin{figure}[!hbtp]
\centering
\includegraphics[bb = 0 0 295 199, scale = 1.0]{\woprfigs 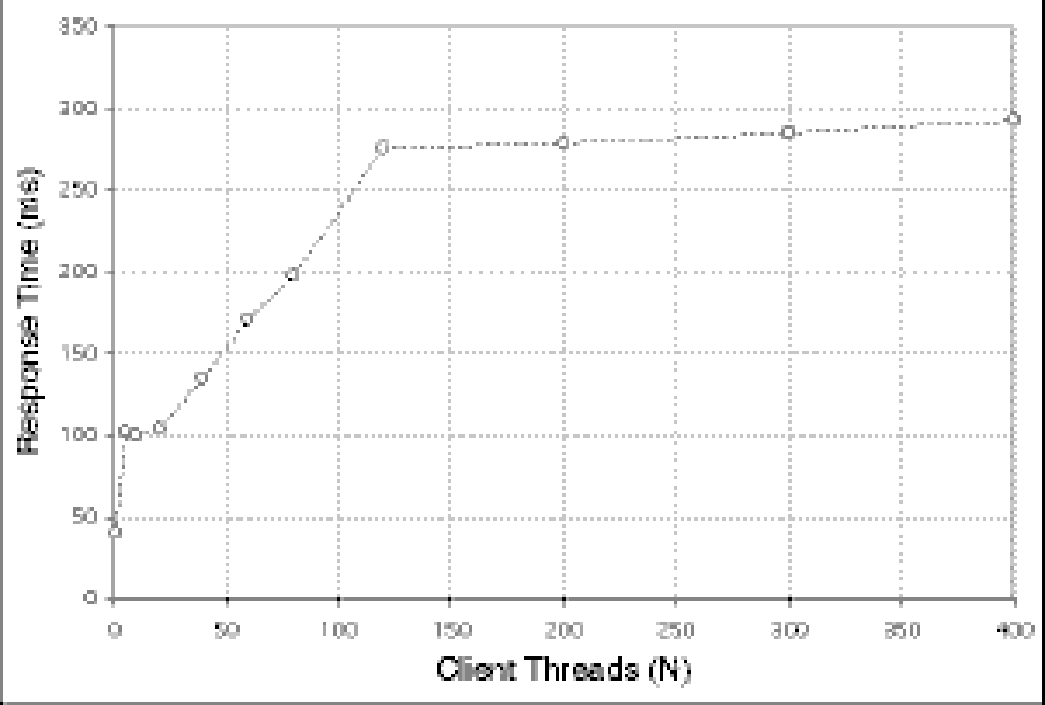} 
\caption{Response time as measured by the WAS tool.} 
\label{fig:strangeIntelR}
\end{figure}

\begin{figure}[!hbtp]
\centering
\includegraphics[bb = 0 0 295 199, scale = 1.0]{\woprfigs 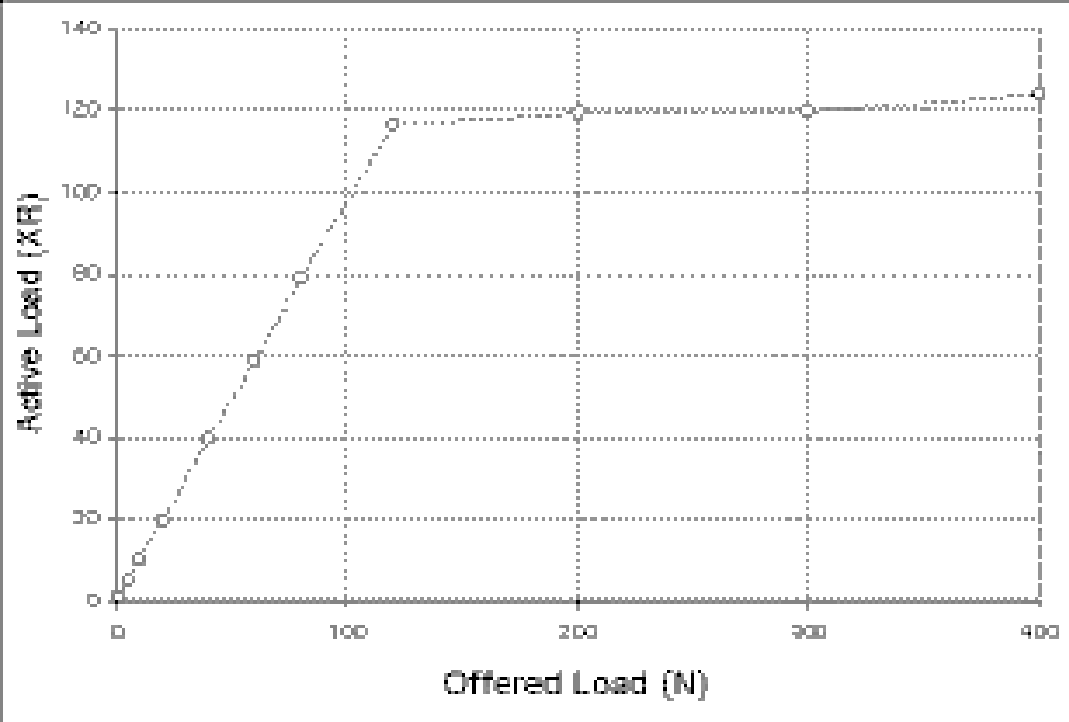} 
\caption{Plot of $N_{run}$ determined by applying LittleÕs law
to the data in Table~\ref{tab:WASthreads}.} 
\label{fig:strangeIntelQ}
\end{figure}

The right approach to analyzing sublinear response times has been presented
by~\citet{Buch}. 
\begin{figure}[!hbtp]
\centering
\includegraphics[bb = 0 0 295 199, scale = 1.0]{\woprfigs 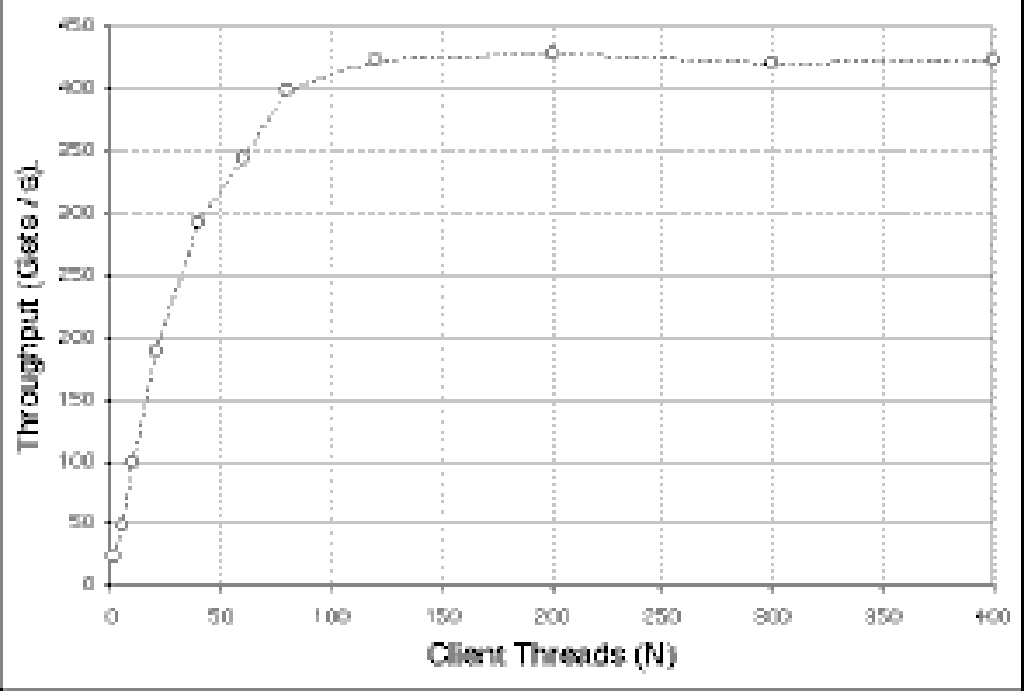} 
\caption{Wintel WAS Throughput.} 
\label{fig:strangeIntelX}
\end{figure}
The context for their measurements is a three-tier e-business application
comprising:
\begin{enumerate}
\item Web services
\item Application services 
\item Database backend
\end{enumerate}
and using Microsoft's Web Application Stress (WAS) tool as the test
harness. The measured throughput in Fig.~\ref{fig:strangeIntelX} exhibits
saturation in the range $100 < N_{was} < 150$ clients. The corresponding
response time data in Fig.~\ref{fig:strangeIntelR} (measured in
milliseconds) exhibits sublinear behavior of the type discussed in
Sects.~\ref{sec:httpd} and~\ref{sec:javaperf}.
In Table~\ref{tab:WASthreads}, $N_{was}$ is the number of client threads
that are assumed to be running. The number of threads that are actually
executing can be determined from the WAS data using Little's
law~\citep{njgBOOK04} in the form $N_{run} = X_{was} \times R_{was}$ (with
the timebase converted to seconds).
\begin{table}[!hbtp]
\caption{} \label{tab:WASthreads}
\begin{tabular}{r @{\hspace{1cm}} r @{\hspace{1cm}} r @{\hspace{1cm}}
r @{\hspace{1cm}} r @{\hspace{1cm}} r}
\hline
\multicolumn{1}{c}{$N_{was}$} &	\multicolumn{1}{c}{$X_{was}$}	& \multicolumn{1}{c}{$R_{was}$} &	\multicolumn{1}{c}{$N_{run}$} & \multicolumn{1}{c}{$N_{idle}$} \\
\hline
1	&		24	&	40	&	0.96	&	0.04 \\
5	&		48	&	102	&	4.90	&	0.10 \\
10	&		99	&	100	&	9.90	&	0.10 \\
... &     ...   &   ... &  ...      & ....  \\
120	&		423	&	276	&	116.75	&	3.25 \\
200	&		428	&	279	& \bf 119.41	&	80.59 \\
300	&		420	&	285	& \bf 119.70	&	180.30 \\
400	&		423	&	293	& \bf 123.94	&	276.06 \\
\hline
\end{tabular}
\end{table}
We see immediately in the fourth column of Table~\ref{tab:WASthreads} that
no more than 120 threads (shown in bold) are ever actually running
(Fig.~\ref{fig:strangeIntelQ}) on the client CPU even though up to 400
client processes have been initiated.
In fact, there are $N_{idle} = N_{was} - N_{run}$ threads that remain idle
in the pool.

This throttling by the size of the client thread pool shows up in the
response data of Fig.~\ref{fig:strangeIntelR} and also accounts for the
sublinearity discussed in Sects.~\ref{sec:httpd} and~\ref{sec:javaperf}.
The complete analysis of this and similar results are presented
in~\citep{njgBOOK04}. Because the other authors were apparently unaware of
the major performance model expressed in Little's law, they failed to
realize that their benchmark was broken.

\section{Conclusion}
Performance data is not divine, even in tightly controlled benchmark
environments. Misinterpreting performance measurements is a common source
of problems---some of which may not appear until the application has been
deployed. One needs a conceptual framework within which to interpret all
performance measurements. I have attempted to demonstrate by example how
simple performance models can fulfill this role. These models provide a way
of expressing our performance expectations. In each of the cited cases, it
is noteworthy that when the data did not meet the expectations set by its
respective performance model, it was not the model that needed to be
corrected but the data. The corollary, having some expectations is better
than having no expectations, follows immediately.

\bibliography{njgWOPR2}
\bibliographystyle{apalike}

\end{document}